\documentclass[10pt]{article}%
\usepackage{amsmath}
\usepackage{amsfonts}
\usepackage{amssymb}
\usepackage{graphicx}%
\usepackage{xcolor}
\providecommand{\U}[1]{\protect\rule{.1in}{.1in}}
\providecommand{\U}[1]{\protect\rule{.1in}{.1in}}
\begin{document}

\begin{center}
{\Large \textbf{Time-Varying Spectrum of the Random String\footnote{It is a
pleasure to dedicate this paper to Prof. Igor Jex, who has made fundamental
contributions to quantum dynamics and the interface with classical mechanics.
In that spirit, this paper extends the results of van Lear Jr. and Uhlenbeck
on the random string to the nonstationary case using the Wigner distribution.}%
}}

\bigskip

\textbf{{L. Galleani\textsuperscript{a} and L.
Cohen\textsuperscript{b,}\footnote{Corresponding author. Email:
leon.cohen@hunter.cuny.edu}} }

\bigskip\textsuperscript{a}Department of Electronics and Telecommunications,
Politecnico di Torino, Corso Duca degli Abruzzi 24, 10129 Torino, Italy

\textsuperscript{b}Department of Physics, Hunter College of the City
University of New York, 695 Park Ave., New York, \newline NY 10065, USA
\end{center}

\begin{center}
\textcolor{red}{\bf{Published in Physica Scripta - Citation data:\\L Galleani and L Cohen 2023 Phys. Scr. 98 014004}\\
DOI 10.1088/1402-4896/aca79e}
\end{center}

\vspace{1em}

\bigskip

\noindent\textbf{{Abstract:}} We consider the response of a finite string to
white noise and obtain the exact time-dependent spectrum. The complete exact
solution is obtained, that is, both the transient and steady-state solution.
To define the time-varying spectrum we ensemble average the Wigner
distribution. We obtain the exact solution by transforming the differential
equation for the string into the phase space differential equation of time and
frequency and solve it directly. We also obtain the exact solution by an
impulse response method which gives a different form of the solution. Also, we
obtain the time-dependent variance of the process at each position. Limiting
cases for small and large times are obtained. As a special case we obtain the
results of van Lear Jr. and Uhlenbeck and Lyon. A numerical example is given
and the results plotted.

\bigskip

\bigskip

\section{Introduction}

Some time after the classical theory of Brownian motion of particles was
developed \cite{hist}, Ornstein \cite{orn} and van Lear Jr. and Uhlenbeck
\cite{lear} considered the application of Brownian motion theory
\cite{uhle2,wang} to extended bodies and in particular to the finite string.
Since that time the random finite string has become a model problem for
studying extended bodies acted on by random forces. Previous work involved the
steady-state spectrum solution and that was generally achieved by starting the
system at minus infinity wherein, therefore, an infinite time has gone by. In
this paper we consider the random string problem and start at a finite time
\cite{lear,lyon}. We obtain the complete solution to the time-varying
spectrum. To define the time-dependent spectrum we use the Wigner spectrum and
obtain the exact solution.

This paper is organized as follows. In the next section we review the string
equation with external forces. Then, we discuss how to handle nonstationary
noise by way of the Wigner spectrum. Subsequently, we obtain the exact
solution for the Wigner spectrum by two different methods and discuss how it
approaches steady state.

\section{String equation with external force}

In preparation for solving the random case, in this section we review the
deterministic string equation and its solution. We use the notation of Van
Lear Jr. and Uhlenbeck and follow their approach. The only notational
difference is that we use $2\mu$ for their $\beta$.

The string equation with damping for unit string density is \cite{lear}%
\begin{equation}
\frac{\partial^{2}s}{\partial t^{2}}+2\mu\frac{\partial s}{\partial
t}=p^{\prime}\frac{\partial s}{\partial x}+p{\frac{\partial^{2}s}{\partial
x^{2}}+}A(x,t) \label{eq8}%
\end{equation}
where $A(x,t)$ is the external applied force which subsequently we take to be
the random force. The approach \cite{lear} to solve Eq. (\ref{eq8}) is to
first solve the homogeneous equation%
\begin{equation}
p{\frac{\partial^{2}s_{H}}{\partial x^{2}}-}\frac{\partial^{2}s_{H}}{\partial
t^{2}}-2\mu\frac{\partial s_{H}}{\partial t}+p^{\prime}\frac{\partial s_{H}%
}{\partial x}=0 \label{eq-2}%
\end{equation}
by separation of variables. This results in a complete set of eigenfunctions
$T_{n}(t)$ and $X_{n}(x)$ which are solutions of
\begin{align}
pX_{n}^{\prime\prime}(x)+p^{\prime}X_{n}^{\prime}(x)  &  =-\lambda_{n}%
X_{n}(x)\label{eq just string}\\
T_{n}^{\prime\prime}(t)+2\mu T_{n}^{\prime}(t)  &  =-\lambda_{n}T_{n}(t)
\end{align}
and where $-\lambda_{n}$ are the eigenvalues. The eigenfunctions, $X_{n}(x)$,
are orthogonal and complete and are assumed to be normalized to
one\footnote{Unless specified, all integrals are from $-\infty$ to $\infty$
and all summations are from $1$ to $\infty$.}%
\begin{equation}
\int X_{n}^{\ast}(x)X_{m}(x)dx=\delta_{nm} \label{eq-16}%
\end{equation}
where $\delta_{nm}$ is the Kronecker delta function.\ The general solution to
the homogeneous equation is then%
\begin{equation}
s_{H}(x,t)=\sum a_{n}T_{n}(t)X_{n}(x)
\end{equation}
where
\begin{equation}
a_{n}=\int s_{H}(x,0)X_{n}^{\ast}(x)dx
\end{equation}

\subsection{Inhomogeneous equation}

To solve the inhomogeneous case, Eq. (\ref{eq8}), one expands the unknown
solution, $s(x,t)$, and the known source term, $A(x,t)$, as%
\begin{align}
s(x,t)  &  =\sum S_{n}(t)X_{n}(x)\label{eq-13}\\
A(x,t)  &  =\sum A_{n}(t)X_{n}(x) \label{eq-14}%
\end{align}
where $A_{n}(t)$ are known and are calculated by way of%
\begin{equation}
A_{n}(t)=\int A(x,t)X_{n}^{\ast}(x)dx \label{eq-15}%
\end{equation}

The equations for the unknowns, $S_{n}(t)$, are obtained by substituting Eq.
(\ref{eq-13})\ and Eq. (\ref{eq-14}) into Eq. (\ref{eq8}) to obtain an
ordinary differential equation for each $S_{n}(t)$
\begin{equation}
S_{n}^{\prime\prime}(t)+2\mu S_{n}^{\prime}(t)+\lambda_{n}S_{n}(t)=A_{n}%
(t)\label{eq25}%
\end{equation}
and which has to be solved for each $S_{n}(t)$. The solution is given by
\cite{lear}%
\begin{equation}
S_{n}(t)=\left[  \frac{\mu S_{n}(0)+S_{n}^{\prime}(0)}{\omega_{n}}\sin
\omega_{n}t+S_{n}(0)\cos\omega_{n}t\right]  e^{-\mu t}+\frac{1}{\omega_{n}%
}\int_{0}^{t}A_{n}(t^{\prime})e^{-\mu(t-t^{\prime})}\sin\omega_{n}%
(t-t^{\prime})dt^{\prime}\label{eq-19}%
\end{equation}
where%
\begin{equation}
\omega_{n}=\sqrt{\lambda_{n}-\mu^{2}}%
\end{equation}
where one must take%
\begin{equation}
\mu^{2}<\lambda_{n}\label{Eq: mu^2 < lambda_n}%
\end{equation}
Once one obtains the $S_{n}(t)$'s the general solution, $s(x,t)$, is given by
Eq. (\ref{eq-13}). Following Van Lear and Uhlenbeck we consider the case of an
initially flat string where $S_{n}(0)=S_{n}^{\prime}(0)=0$. Hence we take%
\begin{equation}
S_{n}(t)=\frac{1}{\omega_{n}}\int_{0}^{t}A_{n}(t^{\prime})e^{-\mu(t-t^{\prime
})}\sin\omega_{n}(t-t^{\prime})dt^{\prime}\label{eq-18}%
\end{equation}

We note that our method allows also to specify non-zero initial conditions of
the string by setting proper values for $S_{n}(0)$ and $S_{n}^{\prime}(0)$ in
Eq. (\ref{eq-19}). Moreover, in Eq. (\ref{eq-19}), we take $A_{n}(t)=0$ for
$t<0$, and, consequently, the initial conditions $S_{n}(0)$ and $S_{n}%
^{\prime}(0)$ hold for $t\geq0$ only.

\subsection{Statistics of $A_{n}(t)$}

In the next section we will be finding the time-frequency spectrum for both
the transient and steady state. We shall need the statistics of $A_{n}(t)$ as
given by Eq. (\ref{eq-15}). In Appendix A we show that\ if we take the driving
force, $A(x,t)$, to be
\begin{equation}
A(x,t)=u(t)\eta(x,t)
\end{equation}
where $u(t)$ is the step function and $\eta(x,t)$ is white Gaussian noise in
space and time, then $A_{n}(t)$ is white Gaussian noise starting at $t=0$,
given by
\begin{equation}
A_{n}(t)=u(t)\xi_{n}(t) \label{eq-20}%
\end{equation}
where $\xi_{n}(t)$ is white Gaussian\ noise for each $n$, with zero mean and
autocorrelation function%
\begin{equation}
E\left[  \xi_{n}(t^{\prime})\xi_{n}(t^{\prime\prime})\right]  =\delta
(t^{\prime}-t^{\prime\prime}) \label{Eq: Autocorrelation White Gaussian noise}%
\end{equation}
where $E$ is the ensemble averaging operator. Note that, since we are dealing
with continuous time, white Gaussian noise has infinite variance, as it can be
seen by setting $t^{\prime}=t^{\prime\prime}$ in Eq.
(\ref{Eq: Autocorrelation White Gaussian noise}).

\section{Time-varying spectrum}

Our aim is\ to calculate the time-varying spectrum for $s(x,t)$ as given by
Eq. (\ref{eq-13}) and Eq. (\ref{eq25}). The Wigner distribution $W_{z}$ for a
deterministic time-dependent function $z(t)$ is \cite{wign}
\begin{equation}
W_{z}(t,\omega)=\frac{1}{2\pi}\int z^{\ast}(t-{{\tfrac{1}{2}}}\tau
)\,z(t+{{\tfrac{1}{2}}}\tau)\,e^{-i\tau\omega}\,d\tau
\end{equation}
The Wigner distribution has found many applications in both time-frequency and
position-momentum space \cite{rev,book,pat,sc,sculzub}. For the time-varying
spectrum we use the Wigner \emph{spectrum, }which, for a stochastic process
$Z(t)$ is defined by \cite{mart1,gal4,book},{%
\begin{equation}
\overline{W}_{Z}(t,\omega)=\frac{1}{2\pi}\int E\left[  Z^{\ast}(t-{{\tfrac
{1}{2}}}\tau)\,Z(t+{{\tfrac{1}{2}}}\tau)\right]  \,e^{-i\tau\omega}%
\,d\tau\label{eq23}%
\end{equation}
}where $E\left[  Z^{\ast}(t_{1})Z(t_{2})\right]  $ is the
two-time\ autocorrelation function. Taking%
\begin{equation}
Z=s(x,t)=\sum S_{n}(t)X_{n}(x) \label{eq24}%
\end{equation}
then substituting Eq. (\ref{eq24})\ into Eq. (\ref{eq23}), taking the ensemble
average and also breaking up the terms into self terms and cross terms we
obtain%
\begin{equation}
\overline{W}(t,\omega;x)=\sum_{n=1}^{\infty}\left\vert X_{n}(x)\right\vert
^{2}\overline{W}_{nn}(t,\omega)+\sum\limits_{n\neq k=1}^{\infty}X_{n}^{\ast
}(x)X_{k}(x)\overline{W}_{nk}(t,\omega) \label{eq36}%
\end{equation}
where
\begin{equation}
\overline{W}_{nk}(t,\omega)=\int E\left[  S_{n}^{\ast}(t-{{\tfrac{1}{2}}}%
\tau)S_{k}(t+{{\tfrac{1}{2}}}\tau)\right]  e^{-i\tau\omega}d\tau\label{eq-30}%
\end{equation}
In Eq. (\ref{eq36}) the prime signifies summation over all $n,k$, except for
$n=k$.

Our aim is to obtain the time-frequency spectrum at position $x$ on the
string, that is $\overline{W}(t,\omega;x)$. We obtain both the transient and
steady-state solution. In Appendix B we show that for white noise the cross
Wigner spectrum is zero%
\begin{equation}
\overline{W}_{nk}(t,\omega)=0\hspace{0.5in}\text{for }n\neq k
\end{equation}
Therefore we only have to concern ourselves with the self terms%
\begin{equation}
\overline{W}_{nn}(t,\omega)=\int E\left[  S_{n}^{\ast}(t-{{\tfrac{1}{2}}}%
\tau)S_{n}(t+{{\tfrac{1}{2}}}\tau)\right]  e^{-i\tau\omega}d\tau\label{eq-27}%
\end{equation}
and the position-dependent Wigner spectrum is then%
\begin{equation}
\overline{W}(t,\omega;x)=\sum_{n=1}^{\infty}\overline{W}_{nn}(t,\omega
)\left\vert X_{n}(x)\right\vert ^{2} \label{eq-37}%
\end{equation}

We note that if we integrate out frequency from $\overline{W}_{nn}(t,\omega)$
we obtain
\begin{equation}
\int\overline{W}_{nn}(t,\omega)d\omega=2\pi E\left[  \left\vert S_{n}%
(t)\right\vert ^{2}\right]
\end{equation}
and the time-dependent position spectrum becomes%
\begin{equation}
\overline{W}(t;x)=2\pi\int\sum_{n=1}^{\infty}E\left[  \left\vert
S_{n}(t)\right\vert ^{2}\right]  \left\vert X_{n}(x)\right\vert ^{2}d\omega
\end{equation}
which is the case considered by van Lear Jr. and Uhlenbeck \cite{lear}.

\subsection{Solution}

There are three approaches we have developed to calculate $\overline{W}%
_{nn}(t,\omega)$. The first is to obtain the differential equation that
$\overline{W}_{nn}(t,\omega)$ satisfies and solve it. The second is to use an
impulse response method. The third is using the solution for $S_{k}$ as given
by Eq. (\ref{eq-19}), calculate $E\left[  S_{n}^{\ast}(t-\frac{1}{2}\tau
)S_{n}(t+\frac{1}{2}\tau)\right]  $ and then obtain $\overline{W}%
_{nn}(t,\omega)$ by way of Eq. (\ref{eq-19}). We have found that the first two
approaches are the most revealing and interesting.

For the first method, consider the Wigner distribution corresponding to
$S_{n}$ as given by Eq. (\ref{eq25}). We have to transform the differential
equation for $S_{n}$ into a differential equation for the Wigner distribution.
A general method to obtain the Wigner distribution that corresponds to the
solution of an ordinary equation has been developed \cite{PhysicsLetters,gal1}%
. When this method is applied to Eq. (\ref{eq25})\ one obtains that the
differential equation for $\overline{W}_{nn}(t,\omega)$ is
\begin{equation}
\left(  \frac{\partial}{\partial t}-p_{1}\right)  \left(  \frac{\partial
}{\partial t}-p_{1}^{\ast}\right)  \left(  \frac{\partial}{\partial t}%
-p_{2}\right)  \left(  \frac{\partial}{\partial t}-p_{2}^{\ast}\right)
\overline{W}_{nn}(t,\omega)=16W_{A_{n}}(t,\omega) \label{eq-38}%
\end{equation}
where $W_{A_{n}}(t,\omega)$ is the Wigner distribution of $A_{n}$%
\begin{equation}
W_{A_{n}}(t,\omega)=\int E\left[  A_{n}^{\ast}(t-{{\tfrac{1}{2}}}\tau
)A_{n}(t+{{\tfrac{1}{2}}}\tau)\right]  e^{-i\tau\omega}d\tau
\end{equation}
and where $p_{1}$, $p_{2}$ are defined by the equation%
\begin{equation}
p_{k}=2\operatorname{Re}\gamma_{k}+2i\left(  \operatorname{Im}\gamma
_{k}-\omega\right)
\end{equation}
and%
\begin{equation}
\gamma_{1,2}=-\mu\pm i\omega_{n}%
\end{equation}
The exact solution to Eq. (\ref{eq-38}) is
\begin{align}
\overline{W}_{nn}(t,\omega)  &  =\frac{1}{2\pi}\frac{u(t)}{(\omega^{2}%
-\lambda_{n})^{2}+4\mu^{2}\omega^{2}}\nonumber\\
&  \times\left[  1-e^{-2\mu t}\left[  \frac{\mu^{2}+(\omega+\omega_{n})^{2}%
}{4\omega\omega_{n}}\left(  \cos2(\omega_{n}-\omega)t+\mu\frac{\sin
2(\omega_{n}-\omega)t}{(\omega_{n}-\omega)}\right)  \right.  \right.
\nonumber\\
&  \left.  \left.  -\frac{\mu^{2}+(\omega-\omega_{n})^{2}}{4\omega\omega_{n}%
}\left(  \cos2(\omega_{n}+\omega)t+\mu\frac{\sin2(\omega_{n}+\omega)t}%
{(\omega_{n}+\omega)}\right)  \right]  \right]  \label{eq27}%
\end{align}

For the second method we consider Eq. (\ref{eq25}) with $A_{n}(t)$ given by
Eq. (\ref{eq-20})
\begin{equation}
S_{n}^{\prime\prime}(t)+2\mu S_{n}^{\prime}(t)+\lambda_{n}S_{n}(t)=A_{n}%
(t)=u(t)\xi_{n}(t)
\end{equation}
One can show that \cite{TransientSpectrum,LangevinHO}%
\begin{equation}
\overline{W}_{nn}(t,\omega)=\int_{0}^{t}W_{h}(t-t^{\prime},\omega)dt^{\prime}
\label{eq-38a}%
\end{equation}
where
\begin{equation}
W_{h}(t,\omega)=\frac{1}{4\omega_{n}^{2}}\left[  W_{L}(t,\omega-\omega
_{n})+W_{L}(t,\omega+\omega_{n})-2W_{L}(t,\omega)\cos2\omega_{n}t\right]
\label{Eq: Wh}%
\end{equation}
and where $W_{L}(t,\omega)$ is given by%
\begin{equation}
W_{L}(t,\omega)=u(t)e^{-2\mu t}\frac{\sin2\omega t}{\pi\omega}%
\end{equation}
Evaluating Eq. (\ref{eq-38a}) results in%
\begin{align}
\overline{W}_{nn}(t,\omega)  &  =\frac{1}{8\pi}\frac{u(t)}{\mu^{2}%
+(\omega-\omega_{n})^{2}}\frac{1}{\omega\omega_{n}}\left[  1-e^{-2\mu
t}\left(  \mu\frac{\sin2(\omega-\omega_{n})t}{\omega-\omega_{n}}+\cos
2(\omega-\omega_{n})t\right)  \right] \nonumber\\
&  -\frac{1}{8\pi}\frac{u(t)}{\mu^{2}+(\omega+\omega_{n})^{2}}\frac{1}%
{\omega\omega_{n}}\left[  1-e^{-2\mu t}\left(  \mu\frac{\sin2(\omega
+\omega_{n})t}{\omega+\omega_{n}}+\cos2(\omega+\omega_{n})t\right)  \right]
\label{Eq: Transient spectrum -- HO}%
\end{align}
While it is not obvious that Eq. (\ref{eq27}) is the same as Eq.
(\ref{Eq: Transient spectrum -- HO})\ they are in fact equal. Both forms are
useful depending on the issues being addressed.

\subsection{Steady-state solution and variance}

Taking the limit $t\rightarrow\infty$ in Eq. (\ref{eq27}) we obtain the
steady-state solution for $\overline{W}_{nn}(t,\omega)$,%
\begin{equation}
\lim_{t\rightarrow\infty}\overline{W}_{nn}(t,\omega)=\frac{1}{2\pi}\frac
{1}{(\omega^{2}-\lambda_{n})^{2}+4\mu^{2}\omega^{2}}%
\end{equation}
Therefore the steady-state solution for the position-dependent Wigner spectrum
is%
\begin{equation}
\overline{W}(\omega;x)=\lim_{t\rightarrow\infty}\overline{W}(t,\omega
;x)=\frac{1}{2\pi}\sum_{n=1}^{\infty}\frac{1}{(\omega^{2}-\lambda_{n}%
)^{2}+4\mu^{2}\omega^{2}}\left\vert X_{n}(x)\right\vert ^{2} \label{eq-42}%
\end{equation}

The variance, $\sigma^{2}(x)$, at position $x$ for the steady-state solution
is given by
\begin{equation}
\sigma^{2}(x)=\int\lim_{t\rightarrow\infty}\overline{W}(t,\omega
;x)d\omega=\frac{1}{2\pi}\sum_{n=1}^{\infty}\int\frac{1}{(\omega^{2}%
-\lambda_{n})^{2}+4\mu^{2}\omega^{2}}d\omega\left\vert X_{n}(x)\right\vert
^{2}%
\end{equation}
The integral evaluates to
\begin{equation}
\int\frac{1}{(\omega^{2}-\lambda_{n})^{2}+4\mu^{2}\omega^{2}}=\frac{\pi}%
{2\mu\lambda_{n}}%
\end{equation}
and therefore%
\begin{equation}
\sigma^{2}(x)=\int\lim_{t\rightarrow\infty}\overline{W}(t,\omega
;x)d\omega=\frac{1}{4\mu}\sum_{n=1}^{\infty}\frac{1}{\lambda_{n}}\left\vert
X_{n}(x)\right\vert ^{2} \label{Eq-45}%
\end{equation}
This is the result obtained by van Lear Jr. and Uhlenbeck \cite{lear}.
However, in the next section we obtain the variance as a function of time.

\section{The time-dependent variance}

We now obtain the variance at each position as a function of time. For a
general zero-mean process, $E[Z(t)]=0$, the variance is given by
\begin{equation}
\sigma_{Z}^{2}(t)=E[Z(t)^{2}]=\int\overline{W}_{Z}(t,\omega)d\omega
\label{Eq: sigma^2_x(t)}%
\end{equation}
For our case we take $Z(t)=S_{n}(t)$. Consider first%
\begin{equation}
\sigma_{S_{n}}^{2}(t)=\int\overline{W}_{nn}(t,\omega)d\omega\label{Eq-47}%
\end{equation}
Substituting Eq. (\ref{Eq: Transient spectrum -- HO}) into Eq. (\ref{Eq-47})
one has
\begin{equation}
\sigma_{S_{n}}^{2}(t)=\frac{1}{4\omega_{n}^{2}}\int_{0}^{t}\int_{-\infty
}^{+\infty}\left[  W_{L}(t-t^{\prime},\omega-\omega_{n})+W_{L}(t-t^{\prime
},\omega+\omega_{n})-2W_{L}(t-t^{\prime},\omega)\cos2\omega_{n}t^{\prime
}\right]  dt^{\prime}d\omega
\end{equation}
Evaluation leads to
\begin{equation}
\sigma_{S_{n}}^{2}(t)=\frac{1}{4\lambda_{n}\mu}-\frac{1}{\omega_{n}^{2}%
}e^{-2\mu t}\left[  \frac{1}{4\mu}+\frac{1}{4\lambda_{n}}\left(  \omega
_{n}\sin2\omega_{n}t-\mu\cos2\omega_{n}t\right)  \right]
\end{equation}
The steady-state solution is given by%
\begin{equation}
\sigma_{S_{n}}^{2}(\infty)=\frac{1}{4\lambda_{n}\mu}%
\end{equation}
For small times we have%
\[
\sigma_{S_{n}}^{2}(t)\sim\frac{1}{4\lambda_{n}\mu}-\frac{1}{\omega_{n}^{2}%
}(1-2\mu t)\left[  \frac{1}{4\mu}+\frac{2\omega_{n}^{2}t}{4\lambda_{n}%
}\right]
\]
which simplifies to
\begin{equation}
\sigma_{S_{n}}^{2}\sim\frac{\mu^{2}}{2\omega_{n}^{2}\lambda_{n}}t
\end{equation}

Therefore the variance as a function of position and time is%
\begin{equation}
\sigma^{2}(t,x)=\sum_{n=1}^{\infty}\left\{  \frac{1}{4\lambda_{n}\mu}-\frac
{1}{\omega_{n}^{2}}e^{-2\mu t}\left[  \frac{1}{4\mu}+\frac{1}{4\lambda_{n}%
}\left(  \omega_{n}\sin2\omega_{n}t-\mu\cos2\omega_{n}t\right)  \right]
\right\}  \left\vert X_{n}(x)\right\vert ^{2}%
\end{equation}
For the steady-state we have%
\begin{equation}
\lim_{t\rightarrow\infty}\sigma^{2}(t,x)=\frac{1}{4\mu}\sum_{n=1}^{\infty
}\frac{1}{\lambda_{n}}\left\vert X_{n}(x)\right\vert ^{2} \label{eq53}%
\end{equation}
which is Eq. (\ref{Eq-45}) and was obtained by van Lear Jr. and Uhlenbeck
\cite{lear}. For small times%
\begin{equation}
\lim_{t\rightarrow0}\sigma^{2}(t,x)\rightarrow t\frac{\mu^{2}}{2}\sum
_{n=1}^{\infty}\frac{1}{\omega_{n}^{2}\lambda_{n}}X_{n}(x)^{2}%
\end{equation}

\section{Undamped case\label{Sect: Undamped case}}

The undamped case is obtained by taking
\begin{equation}
\mu=0
\end{equation}
and Eq. (\ref{eq8}) becomes\
\begin{equation}
\frac{\partial^{2}s}{\partial t^{2}}=p^{\prime}\frac{\partial s}{\partial
x}+p{\frac{\partial^{2}s}{\partial x^{2}}+}A(x,t)
\end{equation}
with%
\begin{equation}
\lambda_{n}=\omega_{n}^{2}%
\end{equation}
The Wigner distribution then becomes
\begin{equation}
\overline{W}_{nn}(t,\omega)=\frac{1}{2\pi}\frac{u(t)}{(\omega^{2}-\omega
_{n}^{2})^{2}}\left(  1-\frac{(\omega+\omega_{n})^{2}}{4\omega\omega_{n}%
}\left(  \cos2(\omega_{n}-\omega)t\right)  -\frac{(\omega-\omega_{n})^{2}%
}{4\omega\omega_{n}}\left(  \cos2(\omega_{n}+\omega)t\right)  \right)
\label{eq58}%
\end{equation}
or%
\begin{equation}
\overline{W}_{nn}(t,\omega)=\frac{1}{8\pi}\frac{u(t)}{\omega\omega_{n}}\left(
\frac{1-\cos2(\omega-\omega_{n})t}{(\omega-\omega_{n})^{2}}-\frac
{1-\cos2(\omega+\omega_{n})t}{(\omega+\omega_{n})^{2}}\right)  \label{eq59}%
\end{equation}

\subsection{ Steady-state solution for the undamped case}

Taking the limit $t\rightarrow\infty$ in Eq. (\ref{eq59}) we obtain the
steady-state solution for $\overline{W}_{nn}(t,\omega)$,%
\begin{equation}
\lim_{t\rightarrow\infty}\overline{W}_{nn}(t,\omega)=\frac{1}{2\pi}\frac
{1}{(\omega^{2}-\omega_{n}^{2})^{2}}%
\end{equation}
Therefore the position-dependent steady-state solution for the Wigner spectrum
is%
\begin{equation}
\lim_{t\rightarrow\infty}\overline{W}(t,\omega;x)=\frac{1}{2\pi}\sum
_{n=1}^{\infty}\frac{1}{(\omega^{2}-\omega_{n}^{2})^{2}}\left\vert
X_{n}(x)\right\vert ^{2}%
\end{equation}

\subsection{Variance}

For the damped case in Eq. (\ref{eq53})\ the time variance is%
\begin{equation}
\sigma^{2}=\int\lim_{t\rightarrow\infty}\overline{W}(t,\omega;x)d\omega
=\frac{1}{4\mu}\sum_{n=1}^{\infty}\frac{1}{\omega_{n}^{2}}\left\vert
X_{n}(x)\right\vert ^{2}%
\end{equation}
which goes to infinity for $\mu\rightarrow0$

\section{Example}

Of particular interest is the case of a string with $p^{\prime}=0$ which was
considered by van Lear Jr. and Uhlenbeck \cite{lear}. Without loss of
generality we take $p=1$ in which case we have for the string equation, Eq.
(\ref{eq8}),%
\begin{equation}
\frac{\partial^{2}s}{\partial t^{2}}+2\mu\frac{\partial s}{\partial t}%
={\frac{\partial^{2}s}{\partial x^{2}}+}A(x,t)
\end{equation}
and for Eq. (\ref{eq just string}) we have
\begin{equation}
X_{n}^{\prime\prime}(x)=-\lambda_{n}X_{n}(x)
\end{equation}
The normalized solutions are given by%
\begin{equation}
X_{n}(x)=\sqrt{\frac{2}{L}}\sin\frac{n\pi}{L}x
\end{equation}
with%
\begin{equation}
\lambda_{n}=\left(  \frac{n\pi}{L}\right)  ^{2}%
\end{equation}
According to Eq. (\ref{Eq: mu^2 < lambda_n}) we must choose $\mu$ so that%
\begin{equation}
\mu^{2}<\lambda_{n}=\left(  \frac{n\pi}{L}\right)  ^{2}%
\end{equation}
\ This must be the case for every $n$. Since the smallest $\lambda_{n}$ is
$\pi/L$ we must take
\begin{equation}
\mu<\frac{\pi}{L}%
\end{equation}
In that case we have that
\begin{equation}
\omega_{n}=\sqrt{\left(  \frac{n\pi}{L}\right)  ^{2}-\mu^{2}}%
\end{equation}
The Wigner spectrum is then given by%
\begin{equation}
\overline{W}(t,\omega;x)=\frac{2}{L}\sum_{n=1}^{\infty}\overline{W}%
_{nn}(t,\omega)\sin^{2}\frac{n\pi}{L}x
\end{equation}
with
\begin{align}
\overline{W}_{nn}(t,\omega)  &  =\frac{1}{8\pi}\frac{u(t)}{\mu^{2}%
+(\omega-\omega_{n})^{2}}\frac{1}{\omega\omega_{n}}\left[  1-e^{-2\mu
t}\left(  \mu\frac{\sin2(\omega-\omega_{n})t}{\omega-\omega_{n}}+\cos
2(\omega-\omega_{n})t\right)  \right] \nonumber\\
&  -\frac{1}{8\pi}\frac{u(t)}{\mu^{2}+(\omega+\omega_{n})^{2}}\frac{1}%
{\omega\omega_{n}}\left[  1-e^{-2\mu t}\left(  \mu\frac{\sin2(\omega
+\omega_{n})t}{\omega+\omega_{n}}+\cos2(\omega+\omega_{n})t\right)  \right]
\end{align}

\subsection{Steady-state solution}

The steady-state solution is when $t\rightarrow\infty$ in which case%
\begin{equation}
\overline{W}_{nn}(t=\infty,\omega)=\frac{1}{2\pi}\frac{1}{(\omega^{2}%
-\lambda_{n})^{2}+4\mu^{2}\omega^{2}}%
\end{equation}
and hence the Wigner spectrum is%
\begin{equation}
\overline{W}(t=\infty,\omega;x)=\frac{1}{2\pi}\sum_{n=1}^{\infty}\frac
{1}{(\omega^{2}-\lambda_{n})^{2}+4\mu^{2}\omega^{2}}\left\vert X_{n}%
(x)\right\vert ^{2} \label{eq84}%
\end{equation}

\subsection{Undamped case}

For the undamped case where we take $\mu=0$
\begin{equation}
\omega_{n}=\sqrt{\lambda_{n}}=\frac{n\pi}{L}%
\end{equation}
and the Wigner spectrum is given by%
\begin{equation}
\overline{W}(t,\omega;x)=\frac{2}{L}\sum_{n=1}^{\infty}\overline{W}%
_{nn}(t,\omega)\sin^{2}\frac{n\pi}{L}x
\end{equation}
where now
\begin{equation}
\overline{W}_{nn}(t,\omega)=\frac{1}{8\pi}\frac{u(t)}{\omega\omega_{n}}\left(
\frac{1-\cos2(\omega-\omega_{n})t}{(\omega-\omega_{n})^{2}}-\frac
{1-\cos2(\omega+\omega_{n})t}{(\omega+\omega_{n})^{2}}\right)
\end{equation}
Using Eq. (\ref{eq84})\ the steady-state solution is%
\begin{equation}
\overline{W}(\infty,\omega;x)=\frac{1}{\pi L}\sum_{n=1}^{\infty}\frac
{1}{(\omega^{2}-\left(  \frac{n\pi}{L}\right)  ^{2})^{2}}\sin^{2}\frac{n\pi
}{L}x
\end{equation}

\subsection{Numerical example}

Following the example of van Lear and Uhlenbeck \cite{lear} we take $L=1$ and
$\mu=\frac{\pi}{8}$ in which case%

\begin{equation}
X_{n}(x)=\sqrt{2}\sin n\pi x
\end{equation}
with
\begin{equation}
\lambda_{n}=\pi^{2}n^{2}%
\end{equation}
and
\begin{equation}
\omega_{n}=\frac{\pi}{8}\sqrt{64n^{2}-1}%
\end{equation}
The Wigner spectrum is then%
\begin{equation}
\overline{W}(t,\omega;x)=2\sum_{n=1}^{\infty}\overline{W}_{nn}(t,\omega
)\sin^{2}n\pi x \label{eq92}%
\end{equation}
with
\begin{align}
\overline{W}_{nn}(t,\omega)  &  =\frac{8}{\pi}\frac{u(t)}{\pi^{2}%
+64(\omega-\omega_{n})^{2}}\frac{1}{\omega\omega_{n}}\left[  1-e^{-\pi
t/4}\left(  \frac{\pi}{8}\frac{\sin2(\omega-\omega_{n})t}{\omega-\omega_{n}%
}+\cos2(\omega-\omega_{n})t\right)  \right] \nonumber\\
&  -\frac{8}{\pi}\frac{u(t)}{\pi^{2}+64(\omega+\omega_{n})^{2}}\frac{1}%
{\omega\omega_{n}}\left[  1-e^{-\pi t/4}\left(  \frac{\pi}{8}\frac
{\sin2(\omega+\omega_{n})t}{\omega+\omega_{n}}+\cos2(\omega+\omega
_{n})t\right)  \right]
\end{align}
The steady-state solution is when $t\rightarrow\infty$%
\begin{equation}
\overline{W}_{nn}(t=\infty,\omega)=\frac{1}{2\pi}\frac{1}{(\omega^{2}-\pi
^{2}n^{2})^{2}+\pi^{2}\omega^{2}/16}%
\end{equation}
and the Wigner spectrum at infinity is given by%
\begin{equation}
\overline{W}(t=\infty,\omega;x)=\frac{1}{2\pi}\sum_{n=1}^{\infty}\frac
{1}{(\omega^{2}-\pi^{2}n^{2})^{2}+\pi^{2}\omega^{2}/16}\left\vert
X_{n}(x)\right\vert ^{2} \label{eq95}%
\end{equation}

In Fig. 1 we show the Wigner spectrum at position $x=L/4$ as given by Eq.
(\ref{eq92}). At $t=0$ the Wigner spectrum is zero, then the energy increases
for each mode and concentrates on the modal frequencies $\omega_{n}$, for
$n=1,2,3$. After the initial transient, the Wigner spectrum reaches a steady
state, as given by Eq. (\ref{eq95}).

In Fig. 2 we take the position at $x=L/2$. For the range of $\omega$ plotted
one would expect the first four modes as given by Eq. (\ref{eq92}). However,
the second and fourth mode are zero because $\sin^{2}n\pi/2$ is zero when
$n=2$ or $4$. In fact for that value of $x=L/2$ all the even modes are zero.

In Fig. 3 we take the position at $x=L/\sqrt{5}$ and now we see all the modes
for the range of frequencies plotted. Not all of the modes are visible, as
some are very small.

In Fig. 4 we take $x=L/4$ and show the first 12 modes of the Wigner spectrum
$\overline{W}(t,\omega,x)$. The fourth, eighth, and twelfth mode are zero
because $\sin^{2}n\pi/4$ is zero for all of the modes whose $n$ is a multiple
of 4. In Fig. 5 we show the Wigner spectrum at position $x=L/10$. For a given
time, the energy of the higher modes decreases slower than for the cases shown
in Figs. 1-3. Note that the tenth mode is zero because $\sin^{2}n\pi/10$ is
zero. Finally, in Fig. 6 the left plot is the Wigner spectrum at $x=L/4$, and
the right plot is for $x=3L/4$. As expected, the two Wigner spectra are
identical because of the symmetry with respect to $x=L/2$ of the $\sin^{2}$
term in Eq. (\ref{eq92}). This shows the consistency of the
calculation.\begin{figure}[ptb]
\centering
\includegraphics[scale=.55]{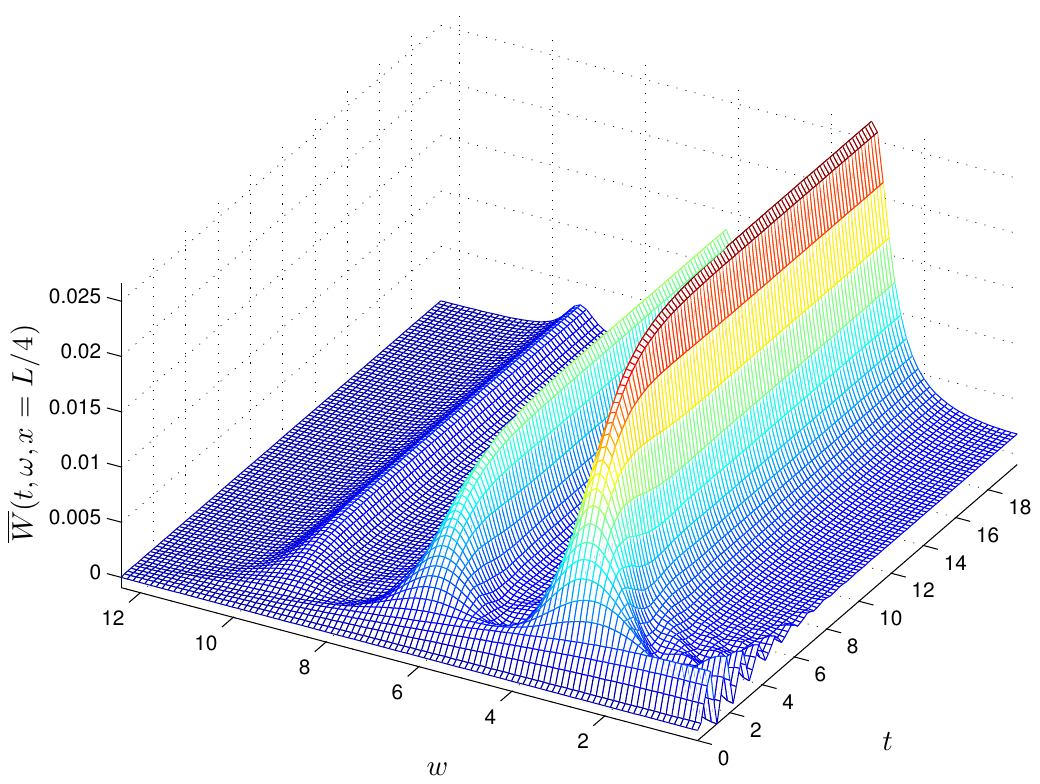}\caption{
The first 3 modes of the Wigner spectrum $\overline{W}(t,\omega,x)$, as given
by Eq. (\ref{eq92}) computed at $x=L/4$. }%
\end{figure}\begin{figure}[ptb]
\centering
\includegraphics[scale=.55]{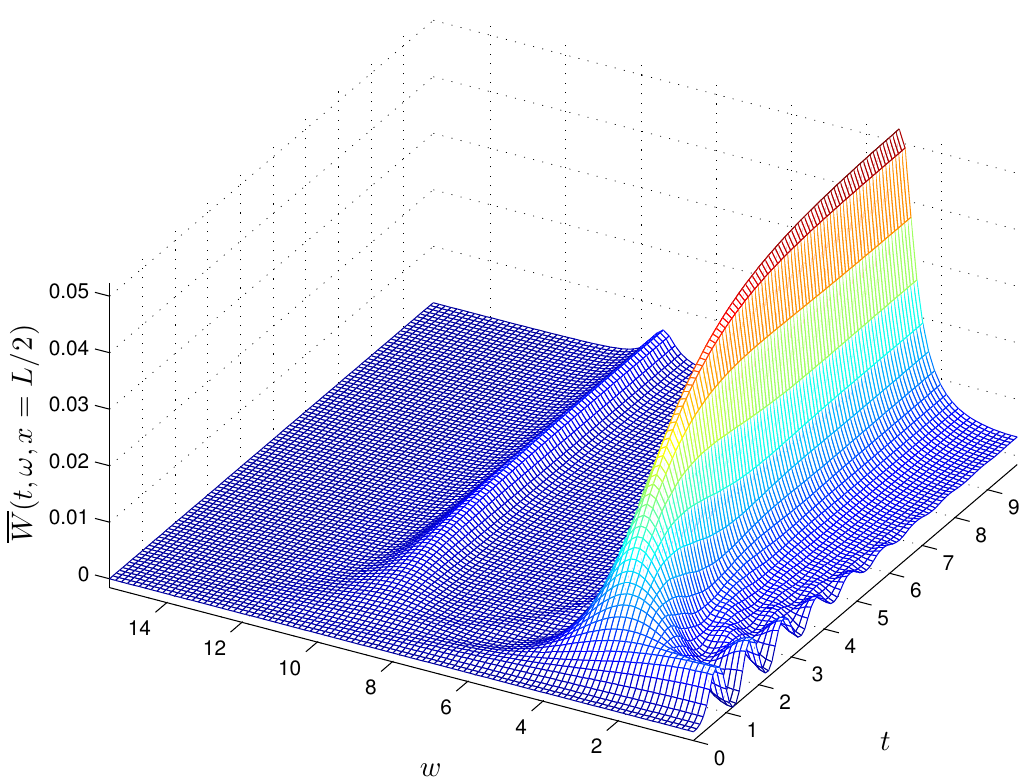}\caption{The
first 4 modes of the Wigner spectrum computed at $x=L/2$. The second and
fourth mode are zero because of the $\sin^{2}$ term in Eq. (\ref{eq92}), which
zeros all of the even modes.}%
\end{figure}\begin{figure}[ptb]
\centering
\includegraphics[scale=.55]{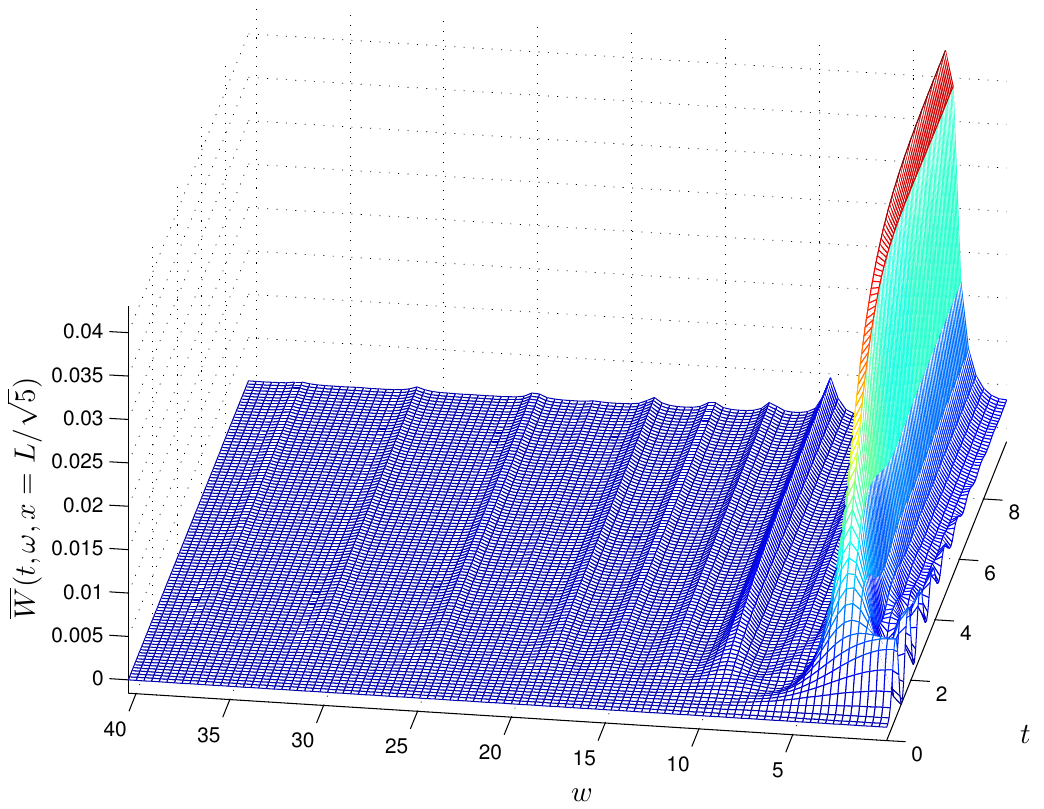}\caption{The
Wigner spectrum at $x=L/\sqrt{5}$. }%
\end{figure}\begin{figure}[ptb]
\centering
\includegraphics[scale=.55]{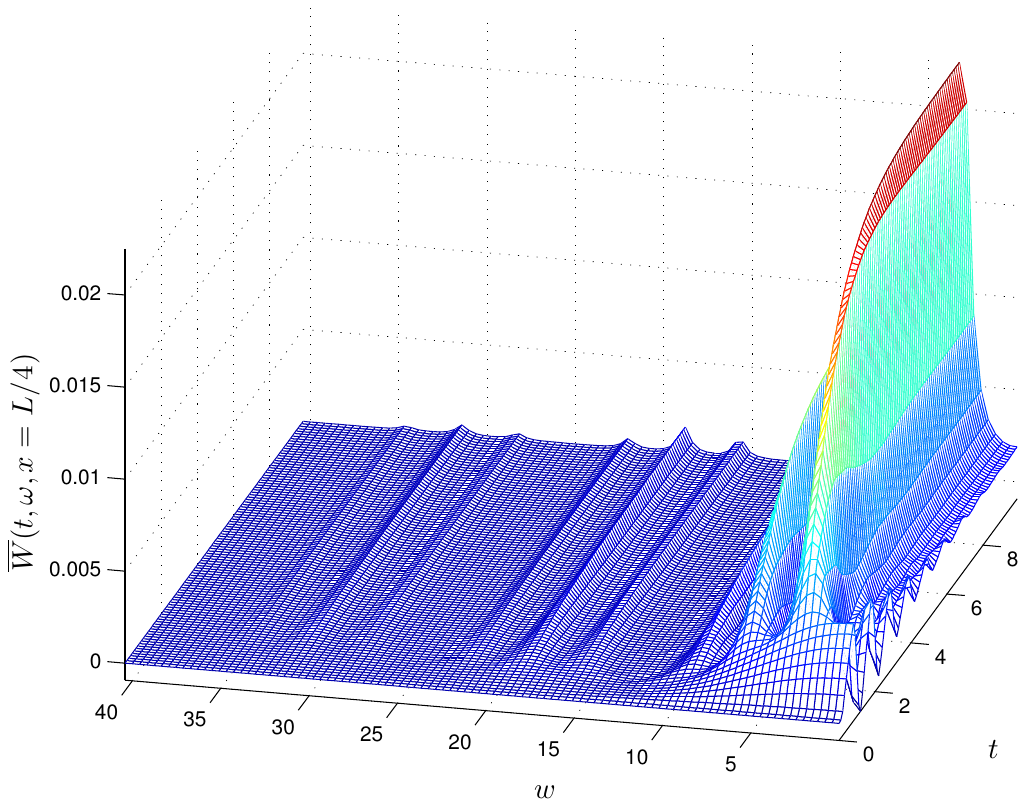}\caption{
The first 12 modes of the Wigner spectrum $\overline{W}(t,\omega,x)$, computed
for $x=L/4$. The fourth, eighth, and twelfth mode are zero because of the
$\sin^{2}$ term in Eq. (\ref{eq92}).}%
\end{figure}\begin{figure}[ptb]
\centering
\includegraphics[scale=.55]{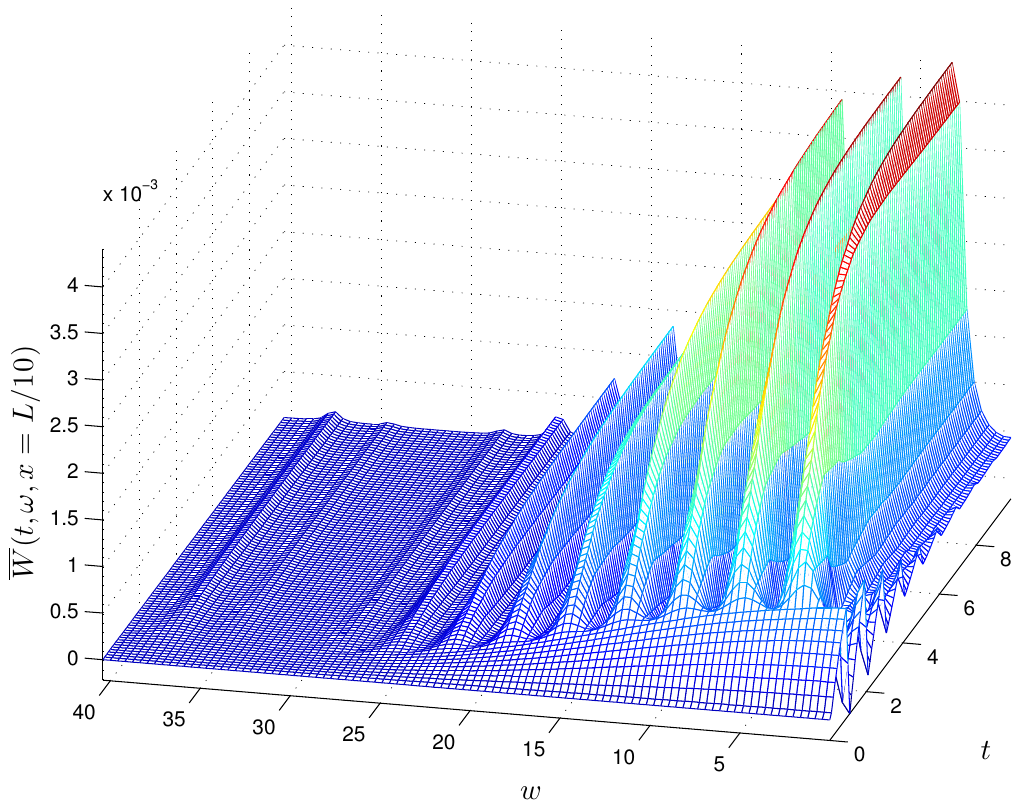}\caption{The
plot shows the first 12 modes of the Wigner spectrum $\overline{W}%
(t,\omega,x)$, computed for $x=L/10$.}%
\end{figure}\begin{figure}[ptb]
\centering
\includegraphics[scale=.55]{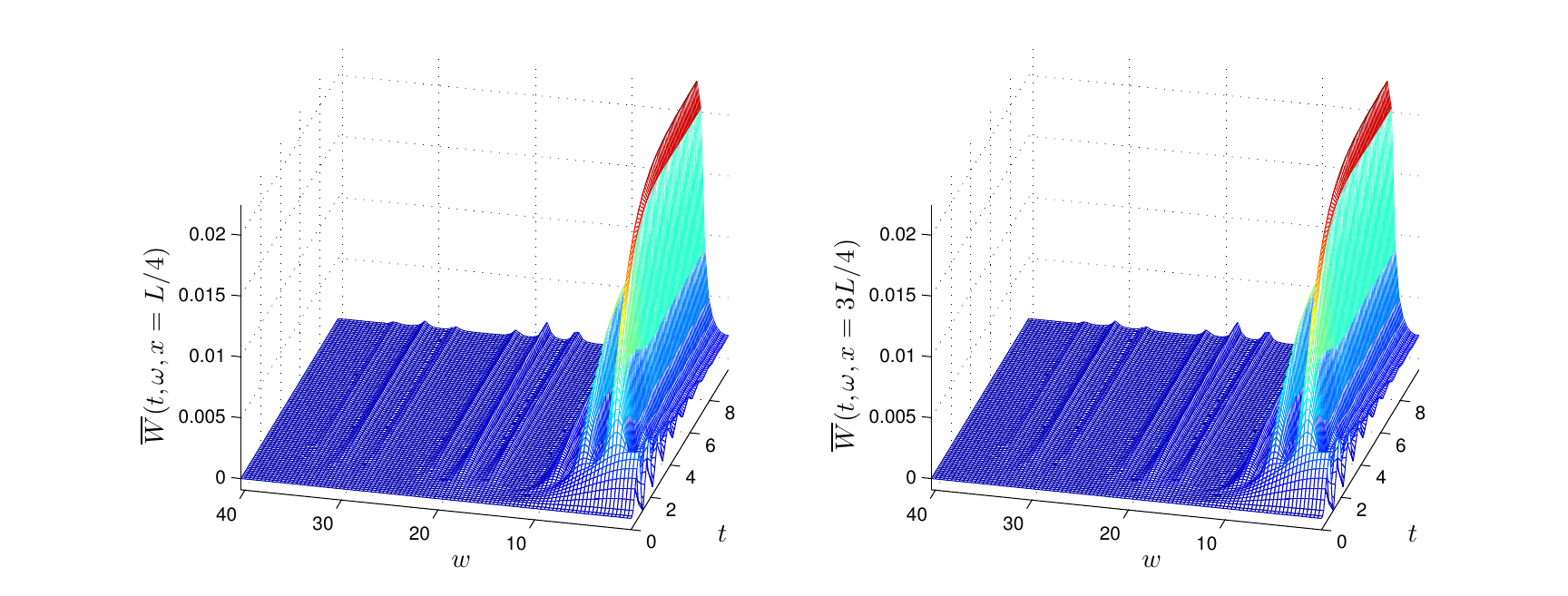}\caption{ The
left plot is the Wigner spectrum at $x=L/4$, whereas the right plot is for
$x=3L/4$. The two spectra are identical because of the symmetry with respect
to $x=L/2$ of the $\sin^{2}$ in Eq. (70).}%
\end{figure}

\section{Summary of results}

For the string equation
\begin{equation}
\frac{\partial^{2}s}{\partial t^{2}}+2\mu\frac{\partial s}{\partial
t}=p^{\prime}\frac{\partial s}{\partial x}+p{\frac{\partial^{2}s}{\partial
x^{2}}+}A(x,t)
\end{equation}
where $A(x,t)$ is the external applied random force the solution is%
\begin{equation}
s(x,t)=\sum S_{n}(t)X_{n}(x)
\end{equation}
where $X_{n}$ are the eigenfunctions of the eigenvalue equation%
\begin{align}
pX_{n}^{\prime\prime}(x)+p^{\prime}X_{n}^{\prime}(x)  &  =-\lambda_{n}%
X_{n}(x)\\
T_{n}^{\prime\prime}(t)+2\mu T_{n}^{\prime}(t)  &  =-\lambda_{n}T_{n}(t)
\end{align}
The equation that $S_{n}(t)$ satisfies is
\begin{equation}
S_{n}^{\prime\prime}(t)+2\mu S_{n}^{\prime}(t)+\lambda_{n}S_{n}(t)=A_{n}(t)
\end{equation}
where%
\begin{equation}
A_{n}(t)=\int A(x,t)X_{n}^{\ast}(x)dx
\end{equation}
The exact Wigner spectrum is
\begin{equation}
\overline{W}(t,\omega;x)=\sum_{n=1}^{\infty}\left\vert X_{n}(x)\right\vert
^{2}\overline{W}_{nn}(t,\omega)
\end{equation}
where%
\begin{align}
\overline{W}_{nn}(t,\omega)  &  =\frac{1}{2\pi}\frac{u(t)}{(\omega^{2}%
-\lambda_{n})^{2}+4\mu^{2}\omega^{2}}\nonumber\\
&  \times\left[  1-e^{-2\mu t}\left[  \frac{\mu^{2}+(\omega+\omega_{n})^{2}%
}{4\omega\omega_{n}}\left(  \cos2(\omega_{n}-\omega)t+\mu\frac{\sin
2(\omega_{n}-\omega)t}{(\omega_{n}-\omega)}\right)  \right.  \right.
\nonumber\\
&  \left.  \left.  -\frac{\mu^{2}+(\omega-\omega_{n})^{2}}{4\omega\omega_{n}%
}\left(  \cos2(\omega_{n}+\omega)t+\mu\frac{\sin2(\omega_{n}+\omega)t}%
{(\omega_{n}+\omega)}\right)  \right]  \right]
\end{align}
or%
\begin{align}
\overline{W}_{nn}(t,\omega)  &  =\frac{1}{8\pi}\frac{u(t)}{\mu^{2}%
+(\omega-\omega_{n})^{2}}\frac{1}{\omega\omega_{n}}\left[  1-e^{-2\mu
t}\left(  \mu\frac{\sin2(\omega-\omega_{n})t}{\omega-\omega_{n}}+\cos
2(\omega-\omega_{n})t\right)  \right] \nonumber\\
&  -\frac{1}{8\pi}\frac{u(t)}{\mu^{2}+(\omega+\omega_{n})^{2}}\frac{1}%
{\omega\omega_{n}}\left[  1-e^{-2\mu t}\left(  \mu\frac{\sin2(\omega
+\omega_{n})t}{\omega+\omega_{n}}+\cos2(\omega+\omega_{n})t\right)  \right]
\end{align}
and where%
\begin{equation}
\omega_{n}=\sqrt{\lambda_{n}-\mu^{2}}%
\end{equation}
The steady-state solution is given by%

\begin{equation}
\overline{W}(\omega;x)=\lim_{t\rightarrow\infty}\overline{W}(t,\omega
;x)=\frac{1}{2\pi}\sum_{n=1}^{\infty}\frac{1}{(\omega^{2}-\lambda_{n}%
)^{2}+4\mu^{2}\omega^{2}}\left\vert X_{n}(x)\right\vert ^{2}%
\end{equation}
The variance as a function of position and time is%
\begin{equation}
\sigma^{2}(t,x)=\sum_{n=1}^{\infty}\left\{  \frac{1}{4\lambda_{n}\mu}-\frac
{1}{\omega_{n}^{2}}e^{-2\mu t}\left[  \frac{1}{4\mu}+\frac{1}{4\lambda_{n}%
}\left(  \omega_{n}\sin2\omega_{n}t-\mu\cos2\omega_{n}t\right)  \right]
\right\}  \left\vert X_{n}(x)\right\vert ^{2}%
\end{equation}

The undamped case is obtained by letting $\mu\rightarrow0$ and the explicit
expressions are given in Sect. \ref{Sect: Undamped case}.

\section{Conclusion}

We have obtained the exact time-dependent spectrum for the random string where
the forcing term is white noise. The time spectrum is defined by way of the
ensemble average of the Wigner distribution. Explicit expressions have been
derived which involve the transient and steady-state solution. The limiting
value for time going to infinity gives the standard spectrum and agrees with
the result of van Lear and Uhlenbeck \cite{lear}. The variance at each
position was explicitly given.

\section{Appendix A: The statistics of $A_{n}(t)$}

Following van Lear and G. E. Uhlenbeck, we take the random force $A(x,t)$ to
be white Gaussian noise in space and time starting at $t=0$,%
\begin{equation}
A(x,t)=u(t)\eta(x,t) \label{Eq: A - Transient case}%
\end{equation}
where $u(t)$ is the step function and $\eta(x,t)$ is a white Gaussian noise in
space and time with autocorrelation function%
\begin{equation}
E\left[  \eta(x^{\prime},t^{\prime})\eta(x^{\prime\prime},t^{\prime\prime
})\right]  =\delta(x^{\prime}-x^{\prime\prime})\delta(t^{\prime}%
-t^{\prime\prime})
\end{equation}
Consequently, the autocorrelation function of the random force is given by%
\begin{equation}
E\left[  A(x^{\prime},t^{\prime})A(x^{\prime\prime},t^{\prime\prime})\right]
=u(t^{\prime})\delta(x^{\prime}-x^{\prime\prime})\delta(t^{\prime}%
-t^{\prime\prime})
\end{equation}
where we have used the simplification $u(t^{\prime})u(t^{\prime\prime}%
)\delta(t^{\prime}-t^{\prime\prime})=u(t^{\prime})\delta(t^{\prime}%
-t^{\prime\prime})$.

We now obtain the statistics of $A_{n}(t)$ as given by Eq. (\ref{eq-15}). For
convenience we take $X_{n}(x)$ to be real and write Eq. (\ref{eq-15}) as
\begin{equation}
A(x,t)=\sum A_{n}(t)X_{n}(x)
\end{equation}
Now consider the autocorrelation function of the $A_{n}(t)$ $^{\prime}s$%
\begin{align}
E\left[  A_{n}(t^{\prime})A_{n}(t^{\prime\prime})\right]   &  =E\left[  \int
A(x^{\prime},t^{\prime})X_{n}(x^{\prime})dx^{\prime}\int A(x^{\prime\prime
},t^{\prime\prime})X_{n}(x^{\prime\prime})dx^{\prime\prime}\right] \\
&  =\iint E\left[  A(x^{\prime},t^{\prime})A(x^{\prime\prime},t^{\prime\prime
})\right]  X_{n}(x^{\prime})X_{n}(x^{\prime\prime})dx^{\prime}dx^{\prime
\prime}%
\end{align}
Using Eq. (\ref{Eq: A - Transient case}), and performing the delta function
integrations and also using Eq. (\ref{eq-16}) we obtain that
\begin{equation}
E\left[  A_{n}(t^{\prime})A_{n}(t^{\prime\prime})\right]  =u(t^{\prime}%
)\delta(t^{\prime}-t^{\prime\prime})
\end{equation}
Moreover, since $A(x,t)$ is Gaussian with mean zero, it also follows that
$A_{n}(t)$ is Gaussian with mean zero%
\begin{equation}
E\left[  A_{n}(t)\right]  =0
\end{equation}
Consequently, $A_{n}(t)$ is a white Gaussian noise starting at $t=0$, and we
can write it as%
\begin{equation}
A_{n}(t)=u(t)\xi_{n}(t)
\end{equation}
where $\xi_{n}(t)$ is a white Gaussian\ noise for each $n$, with zero mean and
autocorrelation function%
\begin{equation}
E\left[  \xi_{n}(t^{\prime})\xi_{n}(t^{\prime\prime})\right]  =\delta
(t^{\prime}-t^{\prime\prime})
\end{equation}

\section{Appendix B: Proof of $\overline{W}_{nk}(t,\omega)=0$}

$\overline{W}_{nk}(t,\omega)$ is given by%
\begin{equation}
\overline{W}_{nk}(t,\omega)=\frac{1}{2\pi}\int E\left[  S_{n}\left(
t-{{\tfrac{1}{2}\tau}}\right)  S_{k}\left(  t+{{\tfrac{1}{2}\tau}}\right)
\right]  e^{-i\tau\omega}d\tau
\end{equation}
For zero initial conditions where%
\begin{equation}
S_{n}(t)=\frac{1}{\omega_{n}}\int_{0}^{t}A_{n}(t^{\prime})e^{-\mu(t-t^{\prime
})}\sin\omega_{n}(t-t^{\prime})dt^{\prime}%
\end{equation}
we have
\begin{align}
\overline{W}_{nk}(t,\omega)  &  =\frac{1}{2\pi}\frac{1}{\omega_{n}^{2}}\int
E\left[  \int_{0}^{t-{{\tau/2}}}A_{n}(t^{\prime})e^{-\mu(t-{{\tau/2}%
}-t^{\prime})}\sin\omega_{n}(t-\tau/2-t^{\prime})dt^{\prime}\right. \\
&  \times\left.  \int_{0}^{t+{{\tau/2}}}A_{k}(t^{\prime\prime})e^{-\mu
(t+\tau/2-t^{\prime\prime})}\sin\omega_{n}(t+\tau/2-t^{\prime\prime
})dt^{\prime\prime}\right]  e^{-i\tau\omega}d\tau\\
&  =\frac{1}{2\pi}\frac{1}{\omega_{n}^{2}}\int\int_{0}^{t-{{\tau/2}}}\int%
_{0}^{t+{{\tau/2}}}E\left[  A_{n}(t^{\prime})A_{k}(t^{\prime\prime})\right]
e^{-\mu(t-{{\tau/2}}-t^{\prime})}e^{-\mu(t+\tau/2-t^{\prime\prime})}\\
&  \times\sin\omega_{n}(t-\tau/2-t^{\prime})\sin\omega_{n}(t+\tau
/2-t^{\prime\prime})e^{-i\tau\omega}dt^{\prime}dt^{\prime\prime}d\tau
\end{align}
But
\begin{align}
E\left[  A_{n}(t^{\prime})A_{k}(t^{\prime\prime})\right]   &  =E\left[  \int
A(x^{\prime},t^{\prime})X_{n}(x^{\prime})dx^{\prime}\int A(x^{\prime\prime
},t^{\prime\prime})X_{k}(x^{\prime\prime})dx^{\prime\prime}\right] \\
&  =\int\int E\left[  A(x^{\prime},t^{\prime})A(x^{\prime\prime}%
,t^{\prime\prime})\right]  X_{n}(x^{\prime})X_{k}(x^{\prime\prime})dx^{\prime
}dx^{\prime\prime}\\
&  =\int\int u(t^{\prime})\delta(x^{\prime}-x^{\prime\prime})\delta(t^{\prime
}-t^{\prime\prime})X_{n}(x^{\prime})X_{k}(x^{\prime\prime})dx^{\prime
}dx^{\prime\prime}\\
&  =u(t^{\prime})\delta(t^{\prime}-t^{\prime\prime})\int X_{n}(x^{\prime
})X_{k}(x^{\prime})dx^{\prime}\\
&  =0
\end{align}
and hence
\begin{equation}
\overline{W}_{nk}(t,\omega)=0
\end{equation}

\end{document}